\documentclass[aps,amsmath,boldmath,twocolumn,showpacs]{revtex4-1}
\usepackage{times}
\usepackage{graphicx}
\usepackage{amssymb}
\usepackage{amsmath}
\usepackage{amsfonts}
\usepackage{dcolumn}
\usepackage{braket}
\usepackage{bm}
\usepackage{multirow}
\usepackage[colorlinks=true,
            linkcolor=red,
            urlcolor=blue,
            citecolor=blue]{hyperref}
\usepackage{amstext}
\usepackage{framed}

\begin{document}

\title{Role of local structural distortion in driving ferroelectricity in GdCrO$_3$}
\author{Sudipta Mahana$^1$$^,$$^2$, U. Manju$^3$, Pronoy Nandi$^1$$^,$$^2$, Edmund Welter$^4$, K. R. Priolkar$^5$ and D. Topwal$^1$$^,$$^2$$^,$}

\email {dinesh.topwal@iopb.res.in, dinesh.topwal@gmail.com}

\affiliation{$^1$Institute of physics, Sachivalaya Marg, Bhubaneswar - 751005, India\\
$^2$Homi Bhabha National Institute, Training School Complex, Anushakti Nagar, Mumbai - 400085, India\\
$^3$CSIR -Institute of Minerals and Materials Technology, Bhubaneswar - 751013, India\\
$^4$Deutsches Elektronen-Synchrotron, Hamburg - 22607, Germany\\
$^5$Department of Physics, Goa University, Taleigao Plateau, Goa - 403206, India}

\date{\today}

\begin{abstract}
Temperature dependent synchrotron x-ray diffraction and extended x-ray absorption fine structure (EXAFS) studies were performed to understand the role of structural characteristics in driving the magnetoelectric multiferoic properties of GdCrO$_3$. The results suggest that the distortion in the structure appears to be associated with the off-center displacement of Gd-atoms together with octahedral rotations via displacement of the oxygen ions in GdCrO$_3$. In addition, the magnetic coupling below magnetic transition temperature leads to additional distortion in the system via magnetostriction effect, playing a complementary role in the enhancement of ferroelectric polarization. Further, a comparative EXAFS study of GdCrO$_3$ with a similar system YCrO$_3$ suggests that oxygen environment of Gd in GdCrO$_3$ is different from Y in YCrO$_3$, which resulting in an orthorhombic ${Pna2_1}$ structure in GdCrO$_3$ in contrast to the monoclinic ${P2_1}$ structure in YCrO$_3$.

\end{abstract}
\pacs{ }
\maketitle
\section{Introduction}
 Magnetoelectric multiferroics have drawn great interest in recent years due to their multi-functionality for a wide variety of potential device applications in modern technologies \cite{ortega2015multifunctional,vopson2015fundamentals,ramesh2007multiferroics}. The family of rare-earth chromites ($R$CrO$_3$) has been recognized as promising systems for multiferroicity at reasonably high temperatures \cite{rajeswaran2012field,ghosh2014polar,ghosh2015atypical,indra2016erratum}. But the conflicting observations of the ferroelectric behavior at relatively high temperature and the average centrosymmetric lattice (${Pbnm}$) and magnetic structure ($G$-type) in these systems remained a puzzling issue in this series of compounds \cite{rajeswaran2012field,zhao2017improper}. GdCrO$_3$, a member of $R$CrO$_3$ family, shows magnetic and ferroelectric transitions simultaneously at around 169 K ($T_M$) \cite{rajeswaran2012field} in contrast to YCrO$_3$ having ferroelectric transition at 473 K and a magnetic transition at 140 K \cite{serrao2005biferroic}. In addition, the strength of the electric polarization in GdCrO$_3$ is one order of magnitude less than that of YCrO$_3$ \cite{rajeswaran2012field,serrao2005biferroic}. Our recent report on temperature dependent x-ray diffraction (XRD) studies along with first-principles density-functional-theory calculations showed that GdCrO$_3$ posseses non-centrosymmetric orthorhombic ${Pna2_1}$ structure \cite{mahana2017local}, in contrast to the monoclinic ${P2_1}$ (non-centrosymmetric) structure in YCrO$_3$ as reported earlier \cite{serrao2005biferroic,ramesha2007observation}, which leads to a weaker polar property in GdCrO$_3$ compared to YCrO$_3$. Phonon instability study in the high symmetry cubic perovskite structure showed that there is a strong similarity between GdCrO$_3$ and YCrO$_3$ as the polar distortion in both the systems is associated with the $R$-ion displacement \cite{serrao2005biferroic,mahana2017local,ray2008coupling}. Furthermore, as the ferroelectric instability is very weak in these systems, the local non-centrosymmetry has been suggested to play a crucial role in driving ferroelectricity \cite{serrao2005biferroic,ramesha2007observation}. Thus, it is essential to study the short-range structural order in these systems to understand the origin of their ferroelectric properties. Since extended x-ray absorption fine structure (EXAFS) is a powerful tool for local structure investigations, we performed detailed EXAFS studies in conjuction with temperature dependent XRD to understand the structure of GdCrO$_3$. These studies show that a strong correlation exists between the presence of local distortion and its implication on the ferroelectric ordering and the global structure of the system. Further, we also discuss a comparative EXAFS study of GdCrO$_3$ with the similar chromite system, YCrO$_3$. 

\section{Experimental details}
Polycrystalline samples of GdCrO$_3$ (YCrO$_3$) were prepared by the solid-state synthesis technique using stoichiometric proportions of Gd$_2$O$_3$ (Y$_2$O$_3$) and Cr$_2$O$_3$ and details are described elsewhere \cite{mahana2017complex,mahana2016giant}. Phase purity of the samples were confirmed by powder XRD measurements carried out in a D8 advanced diffractometer equipped with Cu $K$$_\alpha$ radiation. Temperature-dependent XRD measurements were carried out at the XRD1 beamline at the Elettra synchrotron radiation facility using photons with a wavelength of 0.85507 $\AA{}$. Rietveld refinements of the diffraction patterns were performed using the FULLPROF package. Temperature dependent EXAFS measurements were carried out at P-65 beamline at PETRA-III synchrotron source, DESY, Hamburg, Germany. Both incident ($I_0$) and transmitted ($I_t$) photon intensities were recorded simultaneously  using ionization chambers filled with appropriate gases at Gd-$L_3$ (7243 eV) and Cr-$K$ (5989 eV) edges in GdCrO$_3$ and Y-$K$ (17038 eV) and Cr-$K$ edges in YCrO$_3$ . The raw data collected was background subtracted and normalized to extract EXAFS signals through a series of steps using the ATHENA software \cite{ravel2005athena}. Thereafter, the fitting of the EXAFS spectrum with a specific model obtained from basic crystallographic information was carried out using the ARTEMIS software \cite{ravel2005athena}. The software computes the theoretical spectrum from the given model using the ATOMS and
FEFF6 programs \cite{ravel2005athena,RevModPhys.72.621,rehr2009ab}. 
\section{Results and discussion}
\begin{figure}[!ht]
 \centering
 \includegraphics[height=8.5cm,width=8.5cm]{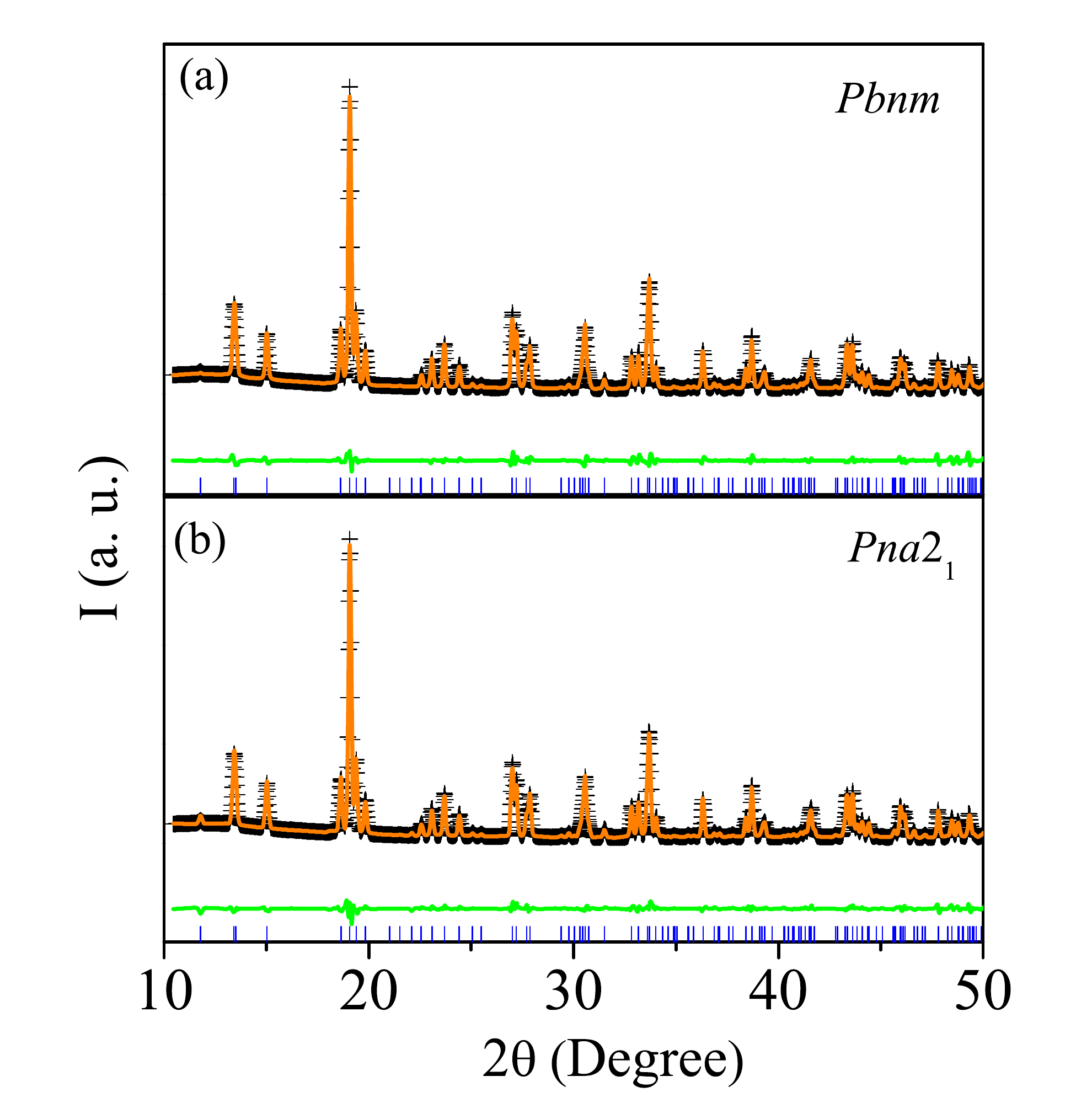}
 \caption{Room temperature x-ray powder diffraction patterns (symbols) of GdCrO$_3$ with corresponding refinement patterns (solid curve) using the (a) ${Pbnm}$ and (b) ${Pna2_1}$ space groups.}
 \label{fig:MT.pdf}
 \end{figure}
\begin{figure*}[!ht]
 \centering
 \includegraphics[height=10.2cm,width=14cm]{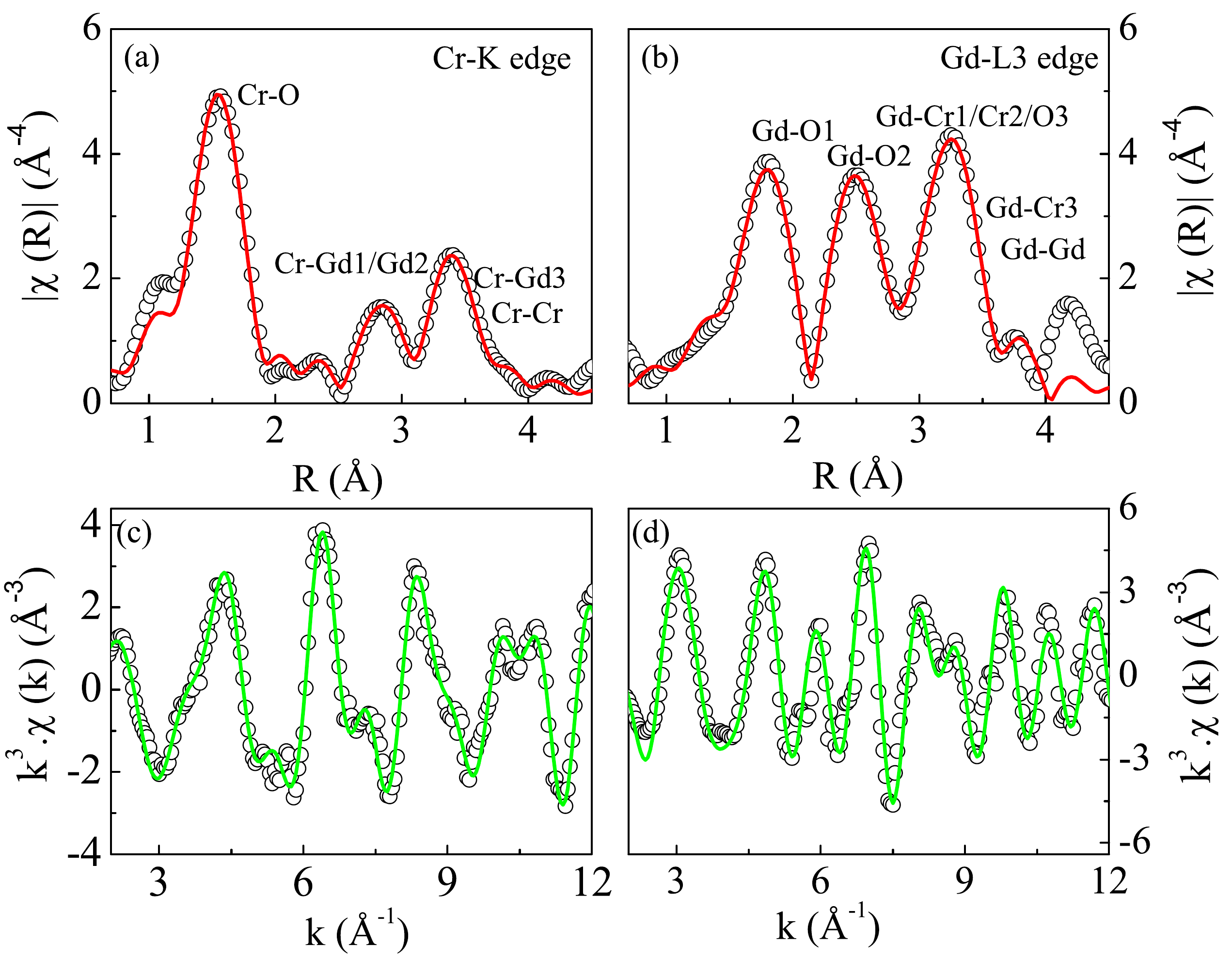}
 \caption{Magnitude of Fourier transforms of  $k^3$-weighted EXAFS data ($\mid$$\chi(R)$$\mid$) at (a) Cr-$K$ edge and (b) Gd-$L_3$ edge acquired at room temperature for GdCrO$_3$, along with corresponding fittings (solid lines). Various contributions in different regions are marked in the figures. Corresponding back-transformed spectra in $k$ space (open circles) along with fitting (solid line) are shown in (c) and (d), respectively. }
 \label{fig:MT.pdf}
 \end{figure*}
\begin{figure}[!ht]
 \centering
 \includegraphics[height=6cm,width=9cm]{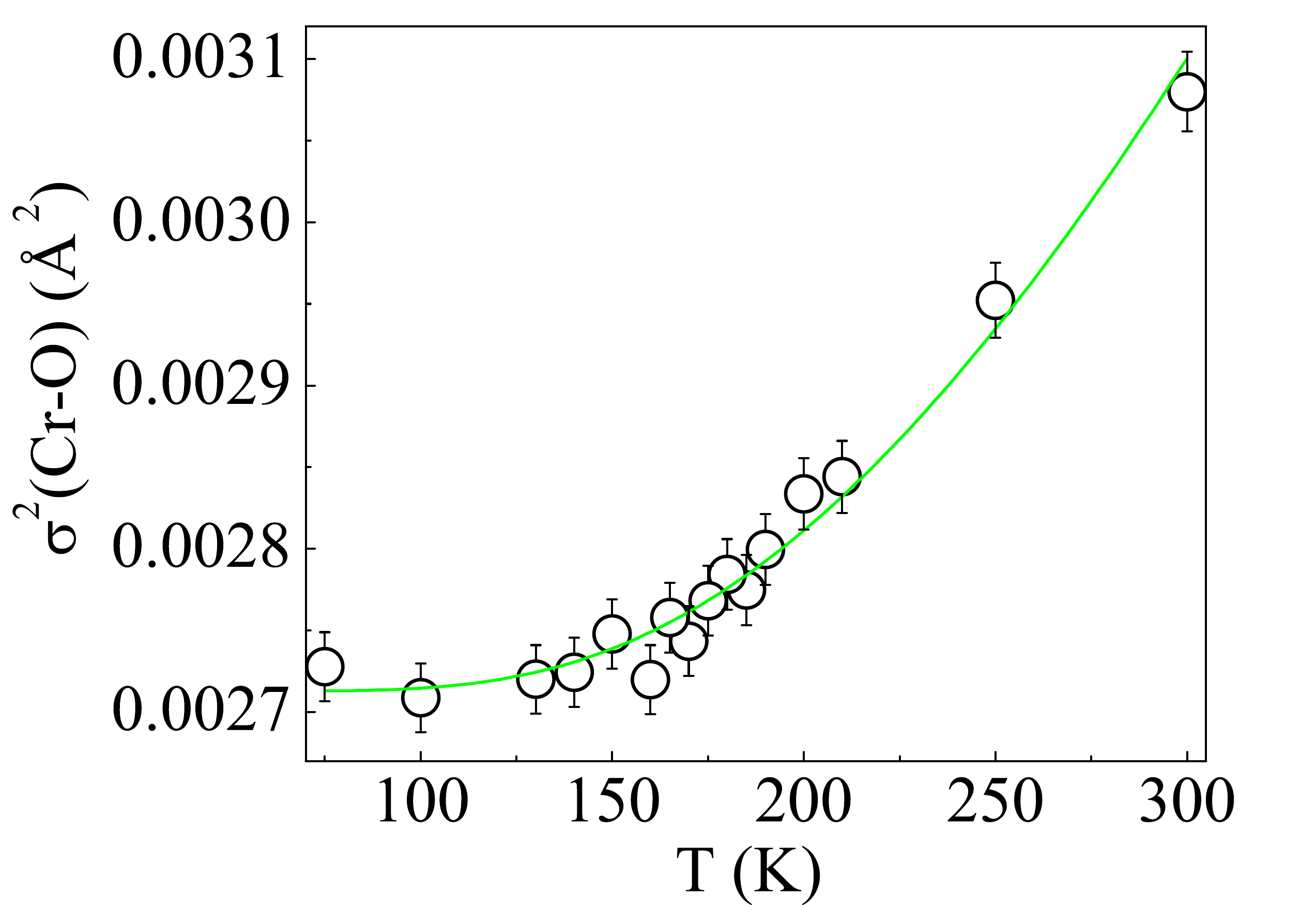}
 \caption{Temperature dependence of MSRD ($\sigma^2$) of  Cr-O bond correlation of GdCrO$_3$. The solid line represents the fitting using Einstein model (Eq. 1). }
 \label{fig:MT.pdf}
 \end{figure}
Rietveld refinement of the XRD patterns of GdCrO$_3$ were performed using both centrosymmetric ${Pbnm}$ and non-centrosymmetric ${Pna2_1}$ space groups throughout the entire (studied) temperature range (100 - 300 K). It was found that both the space groups fit with better goodness-of-fit parameters in the entire range. The XRD pattern acquired at 300 K along with corresponding Rietveld refinement data using ${Pbnm}$ and ${Pna2_1}$ space group are depicted in Fig. 1 (a) and (b), respectively. The reliability parameters obtained for ${Pbnm}$ space group are $R_w$ $\sim$ 0.084, $R_{exp}$ $\sim$ 0.041 and $\chi^2$ $\sim$ 4.16 and the corresponding parameters for ${Pna2_1}$ structure are $R_w$ $\sim$ 0.083, $R_{exp}$ $\sim$ 0.041 and $\chi^2$ $\sim$ 4.37. The similarity in reliability parameters is possibly due to the small structural changes between the two space groups as ${Pna2_1}$ is subgroup of  ${Pbnm}$. The non-centrosymmetric ${Pna2_1}$ structure in GdCrO$_3$  was supported by density-functional theory calculations, favoring polar nature of the system as reported earlier \cite{mahana2017local}.\\

In addition to x-ray diffraction studies, temperature dependent EXAFS measurements were also performed to extract information about the local structure around selected atoms. EXAFS technique is useful to provide valuable information about the structural peculiarities and allows one to verify different structural models. Figures 2 (a) and (b) show the magnitude of Fourier transforms of $k^3$-weighted EXAFS data ($\mid$$\chi(R)$$\mid$)  at Cr-$K$ and Gd-$L_3$ edges, respectively, acquired at room temperature for GdCrO$_3$, along with corresponding fittings superimposed on it. The corresponding back-transformed spectra in $k$ space along with fitting are shown in Figs. 2 (c) and (d), respectively. Various contributions in different regions of the spectra are marked in the figures. The scattering contributions for atomic shells were derived considering both ${Pbnm}$ and ${Pna2_1}$ crystal structures and in either cases the EXAFS data fit well throughout the entire (measured) temperature range having $R$-factor $\sim$ 0.009 for Cr-$K$ edge and $\sim$ 0.007 for Gd-$L_3$ edges. This corroborates that distortions in the structure is very small, in agreement with the XRD results as discussed above. For Cr-$K$ edge, fits were confined to the $k$ range of 3 $<$ $k$ $<$ 12 $\AA{}$$^{-1}$ and $R$-range of 1.2 $<$ $R$ $<$ 4 $\AA{}$. In this region Cr-$K$ EXAFS originates from scattering of photoelectron from the nearest neighbor octahedral oxygens of Cr, three Gd subshells with two, four and two coordination numbers respectively, a Cr-Cr coordination shell with four atoms and some multiple scattering contributions. The fitting of Gd-$L_3$ edge was done in the $k$ range of 3 $<$ $k$ $<$ 12.5  $\AA{}$$^{-1}$ and $R$-range of 1.1 $<$ $R$ $<$ 4 $\AA{}$, to model Gd-O and Gd-Cr distributions. The Gd-O distribution is more complex, consisting of  three subshells with four, two and six oxygens, respectively. The Gd-Cr contribution is also split into three subshells with two, four and two Cr atoms. Additionally, a single Gd-Gd shell is considered with coordination number four. During the fitting procedure, co-ordination number was kept fixed, while bond length and mean square relative displacement (MSRD) [$\sigma^2$ = $<$($r$ - $<$r$>$)$^2$$>$] were used as free parameters. It is seen that MSRD for short (O1) and long (O3) Gd-O bonds are highly correlated, therefore a single $\sigma^2$ for them and separate one for intermediate oxygens (O2). Similarly, two $\sigma^2$ were chosen for Gd-Cr bonds; one for short and long bonds in ${bc}$-plane and another for intermediate bonds. $\sigma^2$ are the most sensitive to modes contributing to radial motions, basically depending only on the local vibrational structure \cite{rehr2000theoretical}. \\
\begin{figure}[!ht]
 \centering
 \includegraphics[height=15cm,width=9cm]{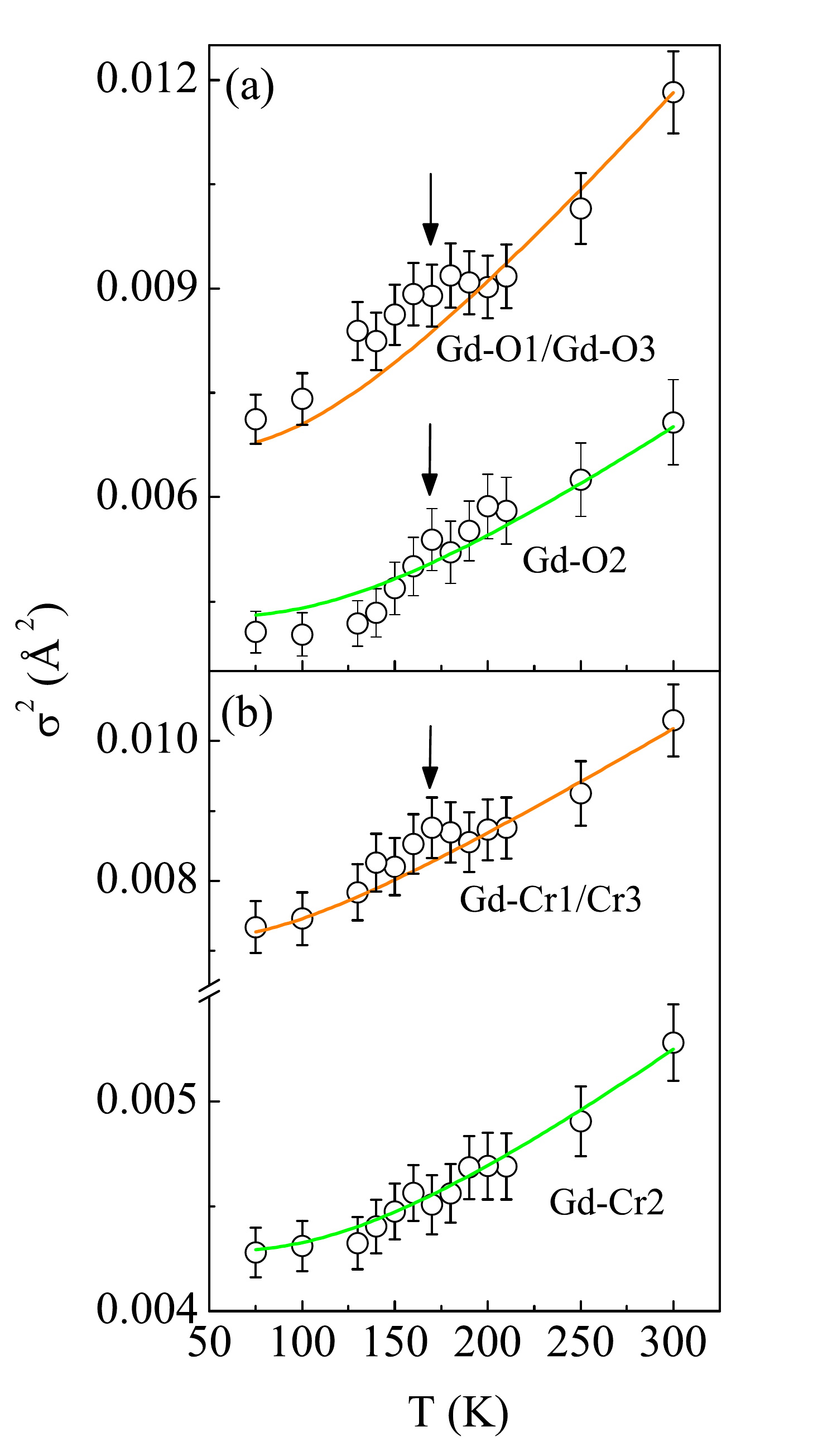}
 \caption{(a) Temperature dependence of MSRD ($\sigma^2$) of  (a) Gd-O bond correlations and (b) Gd-Cr bond correlations in GdCrO$_3$. The solid lines represent the fitting using Einstein model (Eq. 1). The arrows indicate the magnetic/ferroelectric transition temperature.}
 \label{fig:MT.pdf}
 \end{figure}
\begin{figure}[!ht]
 \centering
 \includegraphics[height=11cm,width=8.5cm]{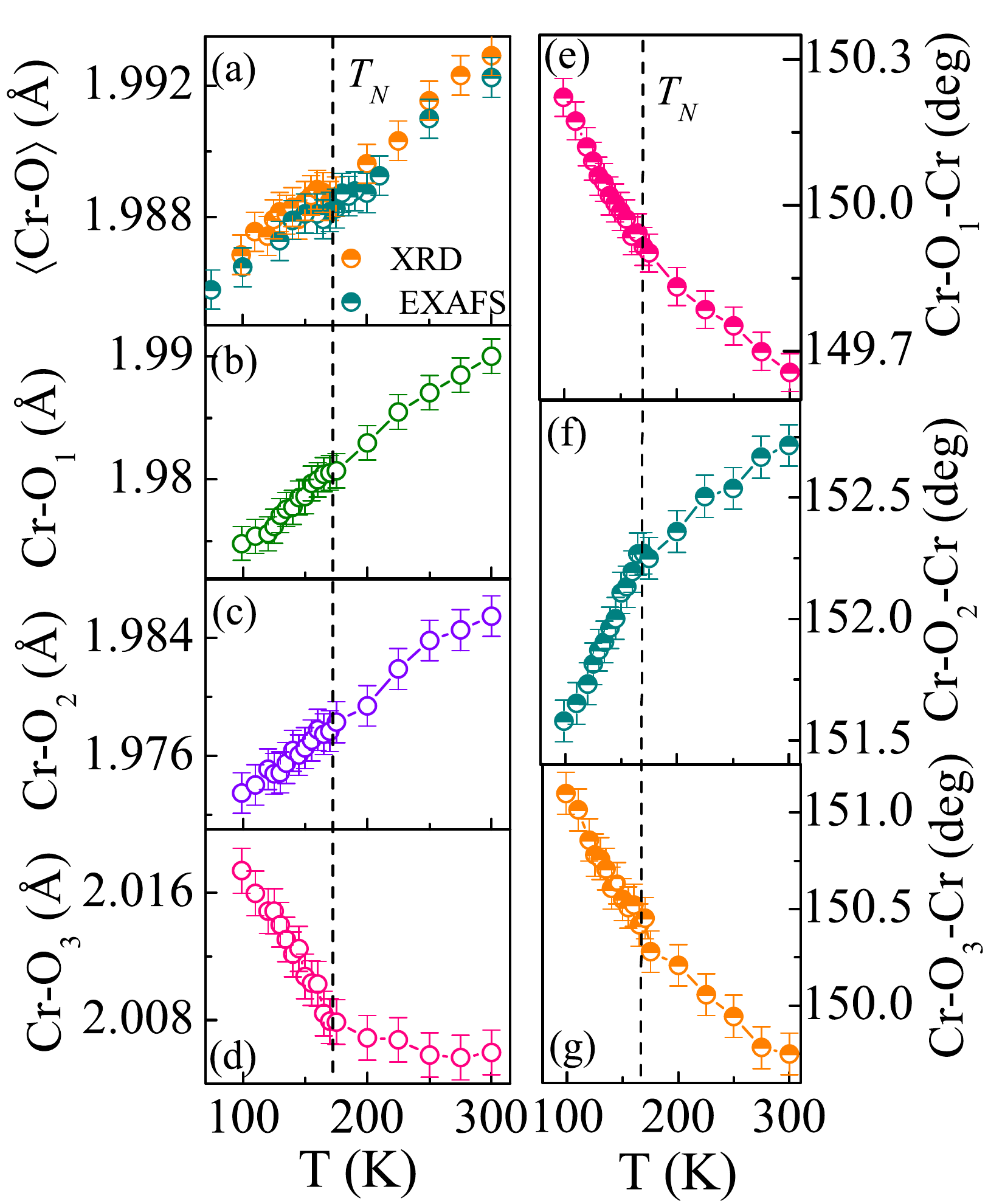}
 \caption{(a) Temperature dependence of average Cr-O atomic bond length in GdCrO$_3$ obtained from EXAFS and XRD analyses. Temperature dependence of bond lengths (Cr-O$_1$/O$_2$/O$_3$) and bond angles (Cr-O$_1$/O$_2$/O$_3$-Cr), obtained from Rietveld refinement of XRD are shown in panels (b)-(g), respectively. Vertical dashed line corresponds to the magnetic/ferroelectric transition temperature.}
 \label{fig:MT.pdf}
 \end{figure}
\begin{figure}[!ht]
 \centering
 \includegraphics[height=10cm,width=12cm]{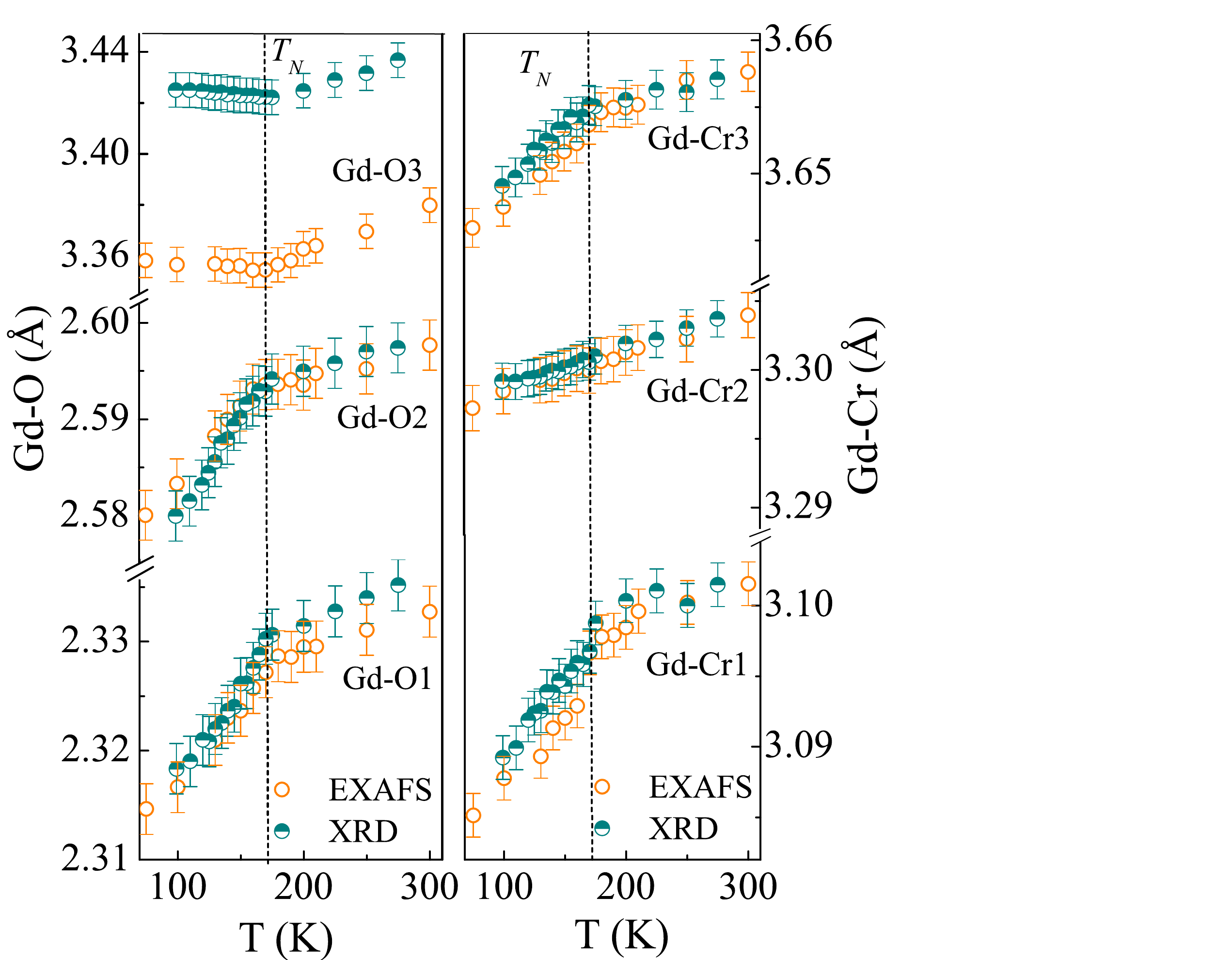}
 \caption{Temperature dependence of  (a) various Gd-O bond lengths and (b) Gd-Cr bond lengths obtained from EXAFS and XRD analyses. Vertical dashed line corresponds to the magnetic/ferroelectric transition temperature.}
 \label{fig:MT.pdf}
 \end{figure}
Temperature dependence of $\sigma^2$ for all the shells were fitted with Einstein model which consider the bond vibrations as harmonic oscillations with a single effective frequency proportional to Einstein temperature, $\theta_E$, as given by the relation \cite{rehr2000theoretical,PhysRevB.45.2447},
\begin{eqnarray}
\sigma^2(T) =  \sigma_0^2 + \bigg(\frac{\hbar^2}{2\mu k_B\theta_E}\bigg) coth \bigg(\frac{\theta_E}{2T}\bigg)
 \end{eqnarray}
where $\sigma_0^2$ is the static contribution, $T$ is in Kelvin and $\mu$ is the reduced mass of the bond pair. The Einstein temperature $\theta_E$ is a measure of the stiffness of the bonds. Fitting of the temperature dependence of $\sigma^2$ using Einstein model for Cr-O bond correlation is shown in Fig. 3. The best fitting yields $\sigma_0^2$ $\sim$ 0.0002 and $\theta_E$ $\sim$ 789 K. The large value of $\theta_E$ indicates the rigidness of the CrO$_6$ octahedra \cite{tyson2007local}. Furthermore, no anomalous change in the $\sigma^2$ is observed below 169 K where both magnetic and ferroelectric ordering are observed. \\

The fitting of temperature dependence of $\sigma^2$ for Gd-O subshells using Eq. (1) are shown in Fig. 4 (a). Fitting yielded $\sigma_0^2$ $\sim$ 0.0013 and $\theta_E$ $\sim$ 335 K for Gd-O1/O3 bond correlations; and $\sigma_0^2$ $\sim$ 0.0003 and $\theta_E$ $\sim$ 412 K for Gd-O2 bond correlation. The relatively low values of $\theta_E$ of Gd-O subshells compared to Cr-O shell suggest that the Gd-O bonds are weaker than Cr-O bonds i.e. GdO$_{12}$ polyhedra are not as rigid as CrO$_6$ octahedra. Furthermore, there is clear deviation of $\sigma_0^2$ from the expected behavior in the region near the magnetic/ferroelectric transition temperature, suggesting the presence of structural anomalies around the transition. The temperature dependent variation of  $\sigma^2$ for Gd-Cr bond correlations along with fitting using Eq. (1) are shown in Fig. 4 (b), giving $\sigma_0^2$ $\sim$ 0.0049 and $\theta_E$ $\sim$ 274 K for Gd-Cr1/Cr3 bonds; and $\sigma_0^2$ $\sim$ 0.0028 and $\theta_E$ $\sim$ 420 K for Gd-Cr2 bond. The $\sigma^2$ for of Gd-Cr1/Cr3 bonds which are in ${bc}$-plane show anomaly around the transition, possibly due to the Gd displacements caused by magnetostriction effect associated with the Gd$^{3+}$-Cr$^{3+}$ interaction \cite{mahana2017local,bhadram2013spin}, whereas Gd-Cr2 bond shows no anomalous behavior.\\

For a further understanding of the structural properties, we have extracted various bond lengths from EXAFS fitting and Rietveld refinement of the XRD pattern and the temperature dependence of average Cr-O bond lengths is depicted in Fig. 5 (a). It is observed that the average bond lengths obtained from both EXAFS and XRD analysis are very close and show no anomalies around the transition temperature. Figure 5 (b-d) represent the individual bond lengths (Cr-O$_1$/O$_2$/O$_3$) extracted from the XRD analysis, in which O$_1$ occupies the apex and O$_2$/O$_3$ occupy the base of CrO$_6$ octahedra. The bond length Cr-O$_1$ contracts with decreasing temperature, while in ${ab}$-plane Cr-O$_2$ bond length decreases and Cr-O$_3$ bond length increases with decreasing temperature. The Cr-O$_3$ bond shows anomalous behavior around the transition indicating magnetostriction effect, consistent with Raman spectroscopy study as reported earlier \cite{mahana2017local}. The temperature dependent variation of the bond angles with temperature are shown in Fig. 5 (e)-(g). With decreasing temperature the axial angle (Cr-O$_1$-Cr) increases and one of the equatorial angles (Cr-O$_2$-Cr) decreases, whereas the other equatorial angle (Cr-O$_3$-Cr) increases. The bond angles also exhibit anomalous behavior near the transition temperature, suggesting that there exist distortions in the octahedra probably associated with the off-center displacements of oxygens via octahedral rotations \cite{ghosh2014polar,ghosh2015atypical}. It is known that the off-center displacement of oxygens, generally in the ${ab}$-plane is the most important factor for the Dzyaloshinskii-Moriya interaction in the system \cite{chiang2011effect,moriya1960anisotropic}.\\

The temperature dependent variation of Gd-O and Gd-Cr bond lengths obtained from EXAFS and XRD analyses are shown in Fig. 6 (a) and (b), respectively.
Except Gd-O3, all the Gd-O and Gd-Cr bonds obtained from EXAFS analysis match reasonably well with that extracted from the Rietveld refinement of XRD. 
Gd-O3 bond obtained from EXAFS fitting is shorter by $\sim$ 0.07$\AA{}$, than that determined from XRD. Discrepancies between EXAFS and diffraction results may arise from either physical or fictitious effects. The systematic errors in the EXAFS data analysis may originate from the correlation between distances and energy scale parameters. On the other hand, discrepancies can also be possible due to the differences between local and long-range structures as observed in various systems like La$_{1-x}$Sr$_x$MnO$_3$ \cite{shibata2003local}, La$_{1-x}$Ca$_x$MnO$_3$ \cite{monesi2005local}, La/PrCoO$_3$ \cite{pandey2006local}, Na$_{0.5}$Bi$_{0.5}$TiO$_3$ \cite{rao2016electric} etc.  In the present system, the nearly equal lengths of Gd-O3 and Gd-Cr bonds may lead to the suppression of Gd-O3 contribution. However, the contraction is observed only for one distance (Gd-O3 bond), while all the other bond lengths show a general agreement between XRD and EXAFS results. Further, though there is a difference in the magnitude of Gd-O3 bond length obtained from XRD and EXAFS analyses, their temperature evolutions are similar. This points towards the true shorter distance of the Gd-O3, which may be possibly due to the deviations of local structure from the average structure. The change in local structure is mainly associated with the Gd-O coordination sphere, whereas the Cr-O coordination remains less affected, suggesting that the structural distortion is dominated by Gd displacements. Further, all the Gd-O bonds and Gd-Cr bonds except Gd-Cr2 show slope changes around the transition temperature. This anomalous behavior can be understood through magnetostriction effect, which plays a role in ferroelectric distortion \cite{mahana2017local,bhadram2013spin}.\\

\begin{figure}[!ht]
 \centering
 \includegraphics[height=14cm,width=8cm]{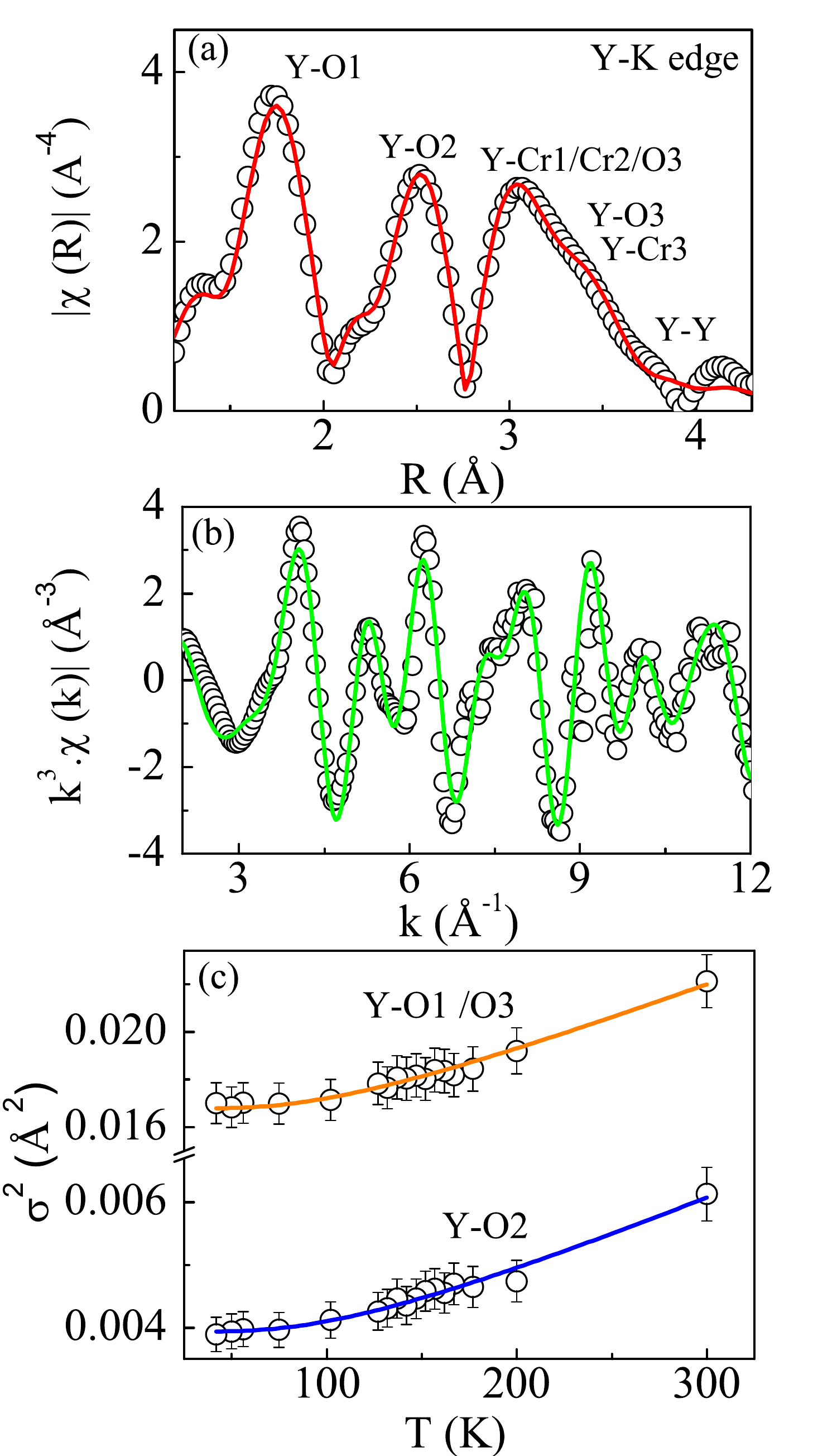}
 \caption{(a) Magnitude of Fourier transform of $k^3$-weighted EXAFS data ($\mid$$\chi(R)$$\mid$) and  (b) back-transformed spectra in $k$ space at Y-$K$ edge acquired at room temperature for YCrO$_3$, along with corresponding fittings (solid lines). Various contributions in different regions are marked in the figure. (c) Temperature dependence of MSRD ($\sigma^2$) of  Y-O bond correlations. The solid lines represent the fitting using Einstein model (Eq. 1).}
 \label{fig:MT.pdf}
 \end{figure}
For comparison the EXAFS measurements were carried out on a similar chromite system, YCrO$_3$. Figures 7 (a) and (b) depict the magnitude of Fourier transform of $k^3$-weighted EXAFS data ($\mid$$\chi(R)$$\mid$) and back-transformed spectra in $k$ space, repectively acquired at Y-$K$ edge at room temperature for YCrO$_3$, along with corresponding fitting superimposed on these. The scattering contributions for atomic shells were derived considering both ${Pbnm}$ crystal structure and  it fits well throughout the entire (measured) temperature range, having $R$-factor $\sim$ 0.01. Fits were confined to the $k$ range of 3 $<$ $k$ $<$ 12.5  $\AA{}$$^{-1}$ and $R$-range of 1.15 $<$ $R$ $<$ 4 $\AA{}$. Here the EXAFS was fitted with three nearest neighbor O shells of four coordination each, three Y-Cr shells with two, four and two atoms, respectively and a single Y shell consisting of four neighbors. A total of five $\sigma^2$ parameters were chosen in a similar way to that used for fitting Gd EXAFS in GdCrO$_3$. The fitting of temperature dependence of $\sigma^2$ for Y-O subshells using Eq. (1) are shown in Fig. 7 (c). From the fitting, it is found $\sigma_0^2$ $\sim$ 0.0018 and $\theta_E$ $\sim$ 330 K for Y-O1/O3 bond distributions; and $\sigma_0^2$ $\sim$ 0.012 and $\theta_E$ $\sim$ 316 K for Y-O2 bond distribution. The relatively low values of $\theta_E$ for  Y-O subshells indicate YO$_{12}$ polyhedra are also not rigid enough like GdO$_{12}$ in GdCrO$_3$. These results suggest a close analogy between GdCrO$_3$ and YCrO$_3$. However, Gd-O environment (three subshells with four, two and six oxygens, respectively) in GdCrO$_3$ is different from Y-O environment (three subshells with four oxygens each) in YCrO$_3$, which leads to orthorhombic ${Pna2_1}$ structure in GdCrO$_3$ in contrast to the monoclinic ${P2_1}$ structure in YCrO$_3$ \cite{mahana2017local,serrao2005biferroic}. The good fitting of EXAFS data using ${Pbnm}$ structure is possibly due to the tiny distortions in these systems, giving rise to very weak polarizations (GdCrO$_3$ $\sim$ 0.7 $\mu$/cm$^2$, YCrO$_3$ $\sim$ 3 $\mu$/cm$^2$) \cite{serrao2005biferroic,rajeswaran2012field} .\\
\begin{figure}[!ht]
 \centering
 \includegraphics[height=8cm,width=8cm]{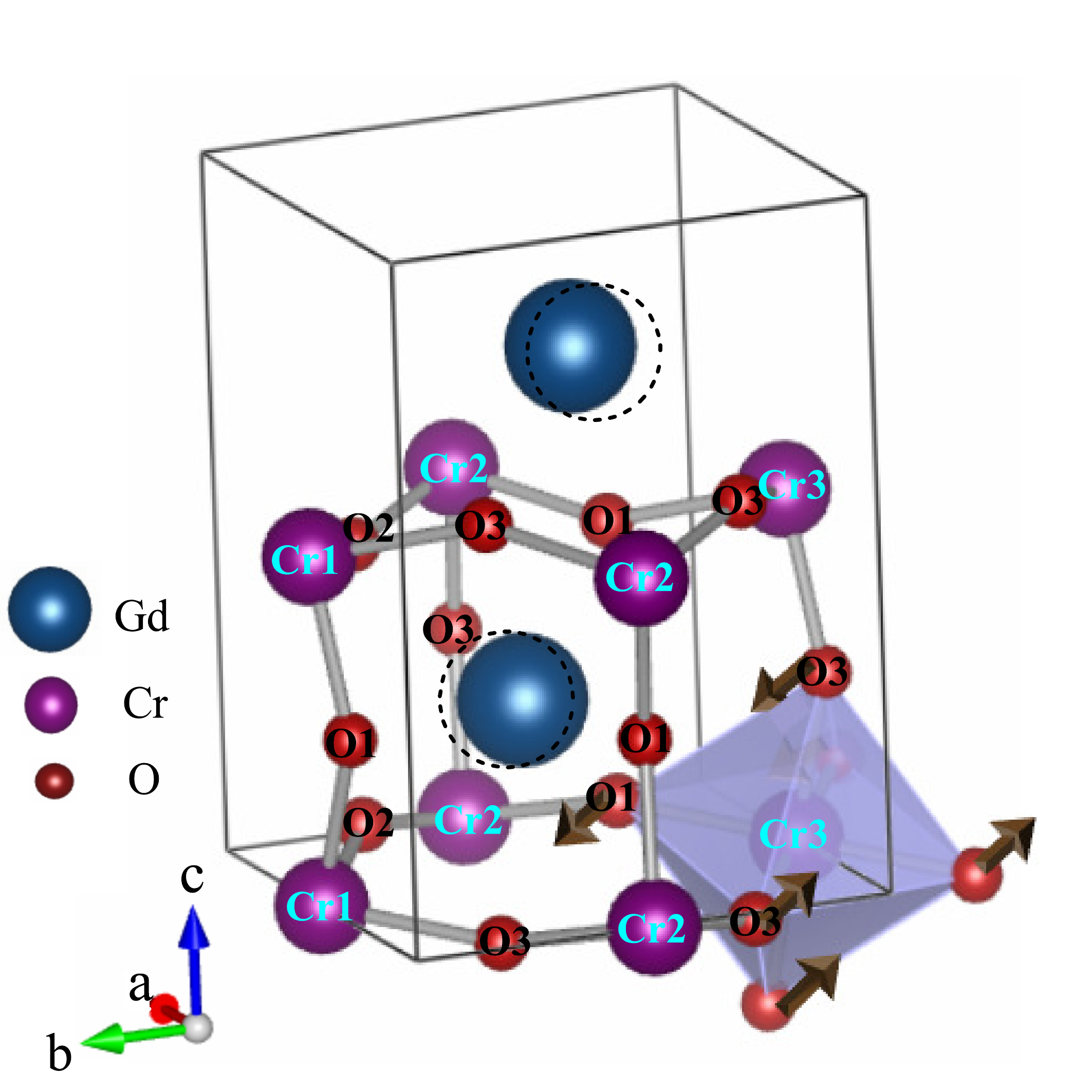}
 \caption{Visualization of displacement of oxygens around CrO$_6$ octahedron (octahedral rotation), as indicated by the arrows and dashed circles represent possible displacement of Gd atoms in GdCrO$_3$.}
 \label{fig:MT.pdf}
 \end{figure}
The distortion in the structure is associated with the off-centering displacement of Gd-atoms together with octahedral rotations via movements of the oxygen ions, which lift certain symmetries of centrosymmetric ${Pbnm}$ structure \cite{ghosh2014polar,ghosh2015atypical} and possibly stabilize a local non-centrosymmeyric ${Pna2_1}$ structure. Figure 8 illustrates a schematic representation of the displacement of oxygens around CrO$_6$ octahedron (octahedral rotation), as indicated by the arrows and dashed circles represent possible displacement of Gd atoms. The distortion is dominated mainly by Gd displacements, indicating that Gd-O bond polarization plays a major role in driving ferroelectricity. Moreover, density-functional theory calculations showed that GdCrO$_3$ prefers ${Pna2_1}$ symmetry as reported earlier and  XRD also gives better fitting with this space group, suggesting that some sort of long-range positional disorder is always present in the system \cite{mahana2017local}. The magnetic coupling leads to extra distortion in the system via magnetostriction effect and plays a complementary role in the enhancement of ferroelectric polarization \cite{mahana2017local,indra2016erratum,bhadram2013spin}. \\

\section{conclusion}
In conclusion, through detailed structural investigations using temperature dependent XRD and EXAFS studies, it is found that CrO$_6$ octahedra in GdCrO$_3$ are rather rigid and the Gd-O bonds are weaker than Cr-O bonds. Octahedral rotations along with Gd displacements lead to non-centrosymmeyric ${Pna2_1}$ structure in GdCrO$_3$. The distortion is dominated by Gd displacements, indicating that Gd-O bond polarization plays a major role in driving ferroelectricity in this system. Furthermore, the magnetoelastic coupling leads to additional distortion in the system resulting in an enhancement of ferroelectric polarization.

\section{Acknowledgment}
 S. M. would like to acknowledge Prof. S. D. Mahanti, Michigan State University, USA for his useful discussions. S.M, U. M., P. N. and D. T. would like to thank the Department of Science and Technology, India for the financial assistance provided to conduct the experiments at Petra III, DESY, Germany under the DST-DESY project. U.M. and D.T. would like to thank International Centre for Theoretical Physics (ICTP), Italy under the ICTP-Elettra users program for the financial support for experiments at Elettra, Italy.

\bibliography{gdcro3}

\bibliographystyle{apsrev4-1}

 \end{document}